\documentclass[english,aps,pra,10pt,tightenlines,twocolumn,superscriptaddress,floatfix]{revtex4-1}
\pdfoutput=1
\usepackage[utf8]{inputenc}
\usepackage[T1]{fontenc}
\usepackage{amsmath}
\usepackage{amssymb}
\usepackage{amsfonts}
\usepackage{bm}
\usepackage{bbm}
\usepackage{epsfig}
\usepackage{grffile}
\usepackage{graphics}
\usepackage{scalefnt}
\usepackage{mathrsfs}

\usepackage[usenames,dvipsnames]{color}
\definecolor{dblue}{rgb}{0,0.1,.6}

\usepackage[colorlinks=true,citecolor=dblue,linkcolor=dblue,urlcolor=dblue]{hyperref}
\usepackage[all]{hypcap}
\newcommand{\Footnote}[1]{\footnote{\unexpanded{#1}}}

\newcommand{\id}{\mathbbm{1}}
\newcommand{\bra}{\langle}
\newcommand{\ket}{\rangle}
\newcommand{\Tr}{\operatorname{Tr}}
\renewcommand{\vec}[1]{{\boldsymbol{#1}}}
\newcommand{\ud}{\mathrm{d}}

\newcommand{\hH}{\hat{H}}
\newcommand{\hL}{\hat{L}}

\newcommand{\dm}{{\hat{\varrho}}}
\newcommand{\vpsi}{\vec{\psi}}
\newcommand{\vphi}{\vec{\phi}}
\newcommand{\ha}{\hat{a}}
\newcommand{\mri}{\mathrm{i}\mkern1mu}
\newcommand{\vk}{\vec{k}}
\newcommand{\vx}{\vec{x}}
\newcommand{\vb}{\vec{b}}

\newcommand{\RR}{\mathbb{R}}
\renewcommand{\Re}{\operatorname{Re}}
\renewcommand{\Im}{\operatorname{Im}}
\newcommand{\groupO} {\operatorname{O}}
\newcommand{\groupU} {\operatorname{U}}

\newcommand{\mc}[1]{\mathcal{#1}}
\newcommand{\pdag}{{\phantom{\dag}}}
\newcommand{\pc}{{\phantom{*}}}

\renewcommand{\L}{\mc{L}}
\newcommand{\LL}{\mathscr{L}}
\newcommand{\D}{\mc{D}}
\newcommand{\eff}{\text{eff}}

\usepackage{amsthm}
\newtheorem*{proposition*}{Proposition}

\renewcommand{\Bmatrix}[1]{\begin{bmatrix}#1\end{bmatrix}}
\newcommand  {\Pmatrix}[1]{\begin{pmatrix}#1\end{pmatrix}}
\newcommand  {\Bsmatrix}[1]{\left[\begin{smallmatrix}#1\end{smallmatrix}\right]}

\newcommand{\duke}  {Department of Physics, Duke University, Durham, North Carolina 27708, USA}
\newcommand{\dqc}   {Duke Quantum Center, Duke University, Durham, North Carolina 27701, USA}

\newcommand{\Title} {Driven-Dissipative Bose-Einstein Condensation and the Upper Critical Dimension}
\newcommand{\Authors}
{
\author{Yikang Zhang}
\affiliation{\duke}
\author{Thomas Barthel}
\affiliation{\duke}
\affiliation{\dqc}
}
\newcommand{\Date} {October 11, 2023}

\begin{document}

\title{\Title}
\Authors
\date{\Date}

\begin{abstract}
Driving and dissipation can stabilize Bose-Einstein condensates. Using Keldysh field theory, we analyze this phenomenon for Markovian systems that can comprise on-site two-particle driving, on-site single-particle and two-particle loss, as well as edge-correlated pumping. Above the upper critical dimension, mean-field theory shows that pumping and two-particle driving induce condensation right at the boundary between the stable and unstable regions of the non-interacting theory. With nonzero two-particle driving, the condensate is gapped. This picture is consistent with the recent observation that, without symmetry constraints beyond invariance under single-particle basis transformations, all gapped quadratic bosonic Liouvillians belong to the same phase. For systems below the upper critical dimension, the edge-correlated pumping penalizes high-momentum fluctuations, rendering the theory renormalizable. We perform the one-loop renormalization group analysis, finding a condensation transition inside the unstable region of the non-interacting theory. Interestingly, its critical behavior is determined by a Wilson-Fisher-like fixed point with universal correlation-length exponent $\nu=0.6$ in three dimensions.
\end{abstract}

\maketitle

\section{Introduction}
Quantum phase transitions \cite{Sondhi1997-69,Vojta2003-66,Sachdev2011} in the context of driven-dissipative systems have received a surge of attention in recent years. Instead of non-analytic changes in zero-temperature states, phase transitions in open many-body systems are characterized by non-analyticities in the non-equilibrium steady state which can arise due to a competition between Hamiltonian terms and environment couplings.

Promising experimental platforms to study such phenomena are cold atoms in optical cavities \cite{Ritsch2013-85,Landig2015-6}, lattices, and tweezers \cite{Bloch2007,Norcia2018-8,Cooper2018-8}, trapped ions \cite{Cirac1995-74,Blatt2008-453,Barreiro2011-470}, Rydberg atoms \cite{Jaksch2000-85,Lukin2001-87,Weimer2010-6,Carr2013-111b,Rao2013-111}, superconducting circuits \cite{Schoelkopf2008-451,Devoret2013-339,Leghtas2013-88}, and  polaritons in circuit-QED or semiconductor-microcavity systems \cite{Hartmann2006-2,Angelakis2007-76,Keeling2007-22,Deng2010-82,Carusotto2013-85,Byrnes2014-10,Hartmann2016-18,Fitzpatrick2017-7, Rodriguez2017-118,Blais2021-93}.

Markovian open quantum systems evolve according to a Lindblad master equation $\partial_t \dm=\L(\dm)$ for the density operator $\dm$ with the Liouvillian
\begin{equation}\label{eq:Lindblad}
	\L(\dm)=-\mri[\hH,\dm]+\sum_{\alpha}\Big(\hL_\alpha^\pdag\dm \hL_\alpha^\dag-\frac{1}{2}\{\hL_\alpha^\dag \hL_\alpha^\pdag,\dm\}\Big),
\end{equation}
where the Lindblad operators $\hL_\alpha$ capture the coupling to the environment \cite{Lindblad1976-48,Gorini1976-17,Breuer2007,Wolf2008-279}. Exact solutions are rare and limited to small systems, quasi-free and quadratic systems \cite{Prosen2008-10,Prosen2010-07,Prosen2010-43,Horstmann2013-87,Guo2017-95,Barthel2021_12}, Yang-Baxter integrable boundary-driven systems \cite{Prosen2011-107b,Prosen2013-15,Karevski2013-110,Prosen2015-48}, or systems with certain dynamical constraints \cite{FossFeig2017-119,Barthel2020_12,McDonald2022-128}. However, the long-distance physics of typical Markovian many-body systems can be analyzed with Keldysh field theory and renormalization group (RG) techniques \cite{Sieberer2016-79,Thompson2023-455,Kamenev2023}. Some examples for this approach can be found in Refs.~\cite{Sieberer2013-110,Sieberer2014-89,Torre2013-87,Maghrebi2016-93,Marino2016-116,Zhang2021-103}.

This work provides a field-theoretical analysis of driven-dissipative Bose-Einstein condensation (BEC) \cite{Diehl2008-4,Diehl2010-105b,Sieberer2013-110,Sieberer2014-89,Carusotto2013-85,Solnyshkov2014-89,Taeuber2014-4,Dunnett2016-93,Zamora2017-7,Walker2018-14,Bloch2022-4,Chen2022-606} above and below the upper critical dimension. Specifically, we consider bosons on a $d$-dimensional cubic lattice, comprising the kinetic energy and an on-site two-particle driving term in the Hamiltonian as well as dissipators for on-site single-particle and two-particle loss and an edge-correlated pumping process,
\begin{subequations}\label{eq:model}
\begin{align}\nonumber
	\L=&-\mri[\hat{H} ,\cdot ] + \frac{\gamma_p}{2}\sum_{\bra i,j\ket} \D[\ha_i^\dag+\ha_j^\dag]\\
	    & + 2d\gamma_l \sum_i \D[\ha_i] + \tilde{u}\sum_i \D[\ha_i^2],  \quad \text{with} \\
	\hat{H}=& -\tilde{J}\sum_{\bra i,j\ket}\ha_i^\dag\ha_j+\sum_i(d\tilde{J}\,\ha_i^\dag\ha_i +G\,\ha^2_i)+H.c.
\end{align}
\end{subequations}
Here, $\D[\hL_\alpha](\dm):=\hL_\alpha\dm \hL_\alpha^\dag-\frac{1}{2}\{\hL_\alpha^\dag \hL_\alpha,\dm\}$ is the dissipator for Lindblad operator $\hL_\alpha$, $\ha_i$ is the bosonic annihilation operator on site $i$ such that $[\ha_i,\ha_j^\dag]=\delta_{i,j}$ and $[\ha_i,\ha_j]=0$, and the sum $\sum_{\bra i,j\ket}$ runs over all lattice edges (pairs of nearest-neighbor sites). The chemical potential has been set to the minimum of the free-boson band.
We could add additional on-site particle pumping terms as well as edge-correlated loss. As long as the latter is weaker than the edge-correlated pumping, these would not change the physics qualitatively.

In Sec.~\ref{sec:action}, we derive the Keldysh action in the continuum limit. According to the mean-field treatment in Sec.~\ref{sec:MFT}, above the upper critical dimension, the condensation transition, as induced by pumping $\gamma_p$ and/or two-particle driving $G$, occurs right at the boundary between the stable and unstable regions of the non-interacting theory \footnote{Physically, unstable open systems correspond to situations where the environment indefinitely pumps energy or particles into the system. Quasi-free fermionic systems are always stable, while quasi-free bosonic systems can be unstable \cite{Barthel2021_12}.}. The tree-level scaling analysis in Sec.~\ref{sec:scaling} yields the upper critical dimension $d_c=4$. In Sec.~\ref{sec:Gaussian}, we discuss the Gaussian approximation for $d>d_c$ finding that, with nonzero two-particle driving $G$, the condensate is gapped and gapless otherwise. In fact, a transition between two gapped phases inside the stable region of the non-interacting theory can be excluded on general grounds by Proposition~5 of Ref.~\cite{Zhang2022-129}. In Sec.~\ref{sec:below}, we carry out the one-loop RG analysis with an $\epsilon$ expansion for $d<d_c$, finding a condensation transition inside the unstable region of the non-interacting theory. The non-equilibrium phase diagram is then determined by a Wilson-Fisher-like fixed point \cite{Wilson1972-28,Wilson1974-12,Altland2010,Taeuber2014-4}. The results are summarized and compared to BEC in closed systems in Sec.~\ref{sec:discussion}. The appendices discuss the experimental realization and provide details for the analytical investigations.

\section{Keldysh action}\label{sec:action}
Similar to procedures for closed systems \cite{Negele1988,Fradkin2013}, we can use a Trotter decomposition of the time evolution operator to write the evolved density operator as
\begin{equation}
	\dm(t)=e^{\L t}\dm(0)=(e^{\L t/N_t})^{N_t}\dm(0).
\end{equation}
Then, inserting resolutions of the identity in terms of coherent states between all factors and taking the trace gives the partition function 
$Z$ as a product of coherent-state matrix elements $\bra\vpsi_+'|e^{\L\delta t}\big[|\vpsi_+\ket\bra\vpsi_-|\big]|\vpsi_-'\ket$ with $\ha_i|\vpsi_\pm\ket=\psi_{\pm,i}|\vpsi_\pm\ket$. Taking the continuous-time limit $N_t\to\infty$, one arrives at \cite{Kamenev2023,Sieberer2016-79}
\begin{subequations}\label{eq:Z}
\begin{align}
	Z&=\int\mathscr{D}[\psi_\pm^\pc,\psi_\pm^*] e^{-S} \quad \text{with Keldysh action}\\
	S&=\int_t\big( \vpsi_+^\dag\partial_t\vpsi_+^\pdag-\vpsi_-^\dag\partial_t\vpsi_-^\pdag -\LL(\vpsi_\pm,\vpsi_\pm^*) \big),
\end{align}
\end{subequations}
where $\LL(\vpsi_\pm,\vpsi_\pm^*)$ are the matrix elements of the Liouvillian. Staring with lattice spacing $a$, we can take the spatial continuum limit by replacing the variables $\psi_{\pm,i}$ with fields $a^{d/2}\psi_\pm(\vx_i)$ and sums $\sum_i$ by integrals $a^{-d}\int_\vx\equiv a^{-d}\int\ud^d x$. For terms that act on lattice edges $\bra i,j\ket$, we express $\psi_\pm(\vx_j)$ in terms of $\psi_\pm(\vx_i)$ and its derivatives up to second order. This step assumes that relevant field configurations are sufficiently smooth, and we will indeed find the gap to close at quasi-momentum $k=0$, such that long-range fluctuations dominate and the gradient expansion captures the relevant physics.
We arrive at the Keldysh action
\begin{align}
\nonumber
	S&=\int_{\vx,t} \Big[ \psi_+^*\partial_t\psi_+^\pc-\psi_-^*\partial_t\psi_-^\pc
	  -2d(\gamma_l\psi_+^\pc\psi_-^* +\gamma_p\psi_+^*\psi_-^\pc) \\ \nonumber
	&\ +d(\gamma_l+\gamma_p) (\psi_+^*\psi_+^\pc +\psi_-^*\psi_-^\pc)\\ \nonumber
	&\ -\frac{a^2\gamma_p}{4}\left(2 \psi_-^\pc\nabla^2\psi_+^*
	                             - \psi_+^*\nabla^2\psi_+^\pc -\psi_-^*\nabla^2\psi_-^\pc \right)\\ \nonumber
	&\ -\mri a^2 \tilde{J}\Big(  \psi_+^*\nabla^2\psi_+^\pc-\psi_-^*\nabla^2\psi_-^\pc \Big)
	  +\mri G(\psi_+^2-\psi_-^2+\text{c.c.})\\
	&\ -\frac{a^d \tilde{u}}{2} \Big(2\psi_+^2\psi_-^{*2}-|\psi_+|^4-|\psi_-|^4\Big) \Big].
	\label{eq:action-pm}
\end{align}
Note that the two-particle driving term in the Hamiltonian breaks the super-particle number conservation such that $[\dm,\sum_i\ha_i^\dag\ha_i]\neq 0$. Hence, for non-zero $G$, the system only has the discrete $\mc{PT}$ symmetry \cite{Bender2018}, while it has a continuous $\groupU(1)$ symmetry under
\begin{equation}\label{eq:U1symmetry}
	\ha_i\mapsto e^{\mri \alpha}\ha_i
	\quad\text{when}\quad
	G=0.
\end{equation}

For the further analysis, it is useful to perform the Keldysh rotation \cite{Kamenev2023} from fields $\psi_\pm$ to
\begin{equation}
	\psi_c:=(\psi_++\psi_-)/{\sqrt{2}},\quad \psi_q:=(\psi_+-\psi_-)/{\sqrt{2}},
\end{equation}
which leads to the action
\begin{align}\nonumber
	&S=\int_{\vx,t} \Big[ \psi_c^*\partial_t\psi_q^\pc+\psi_q^*\partial_t\psi_c^\pc
	  -t_1(\psi_c^*\psi_q^\pc-\psi_q^*\psi_c^\pc)\\ \nonumber
	&\ + t_2\psi_q^*\psi_q^\pc
	   + K_1\left( \psi_c^*\nabla^2\psi_q^\pc -\psi_q^*\nabla^2\psi_c^\pc \right ) + 2K_2 \psi_q^*\nabla^2\psi_q^\pc\\ \nonumber
	&\ -\mri J\left( \psi_c^*\nabla^2\psi_q^\pc+\psi_q^*\nabla^2\psi_c^\pc \right)
	   +2\mri G(\psi_c\psi_q+\psi_c^*\psi_q^*)\\
	&\ +\frac{u}{2}\left(\psi_c^2\psi_c^*\psi_q^*+\psi_q^2\psi_c^*\psi_q^*-\text{c.c.}\right)+2u\psi_c^\pc\psi_c^*\psi_q^\pc\psi_q^*\Big],
	\label{eq:action-cq}
\end{align}
where we have reparametrized the model with
\begin{align}
	&t_1:=d(\gamma_l-\gamma_p),\ \
	t_2:=2d(\gamma_l+\gamma_p)\geq 0,\ \
	\ J:=a^2\tilde{J},\nonumber \\ 
	&K_1=K_2:=a^2 \gamma_p/4\geq 0,\ \ \text{and} \ \
	u:=a^d\tilde{u}\geq 0.
	\label{eq:parameterDef}
\end{align}
We have introduced two separate parameters $K_1$ and $K_2$ which, while they are identical in the original model, will turn out to scale differently in the RG analysis.

\section{Mean-field theory}\label{sec:MFT}
Let us denote the solution of the saddle-point equations
\begin{equation}\label{eq:MF}
	\frac{\delta S}{\delta \psi_c^*}=0, \quad \frac{\delta S}{\delta \psi_q^*}=0
\end{equation}
by $\bar{\psi}_c$, $\bar{\psi}_q$.
Due the conservation of probability, all terms in the action \eqref{eq:action-cq} are at least linear in $\psi_q$ or $\psi_q^*$ \cite{Kamenev2023,Sieberer2016-79}. Hence, the first equation leads to $\bar{\psi}_q=0$. Then, the second equation yields
\begin{equation}\label{eq:MF_q}
	t_1\bar{\psi}_c+2\mri G\bar{\psi}_c^*+\frac{u}{2}\,\rho\bar{\psi}_c=0
	\quad\text{with}\quad
	\rho\equiv |\bar{\psi}_c|^2.
\end{equation}
Multiplying this by $\bar{\psi}_c$ and $\bar{\psi}_c^*$ respectively, we find
\begin{equation}\label{eq:solvingMF}
	\bar{\psi}_c^2 = \frac{-2\mri G\rho}{t_1+\frac{u}{2}\,\rho}
	\quad\text{and}\quad
	\bar{\psi}_c^{*2}=\frac{\mri}{2G}\left(t_1+\frac{u}{2}\,\rho\right)\rho.
\end{equation}
Now, using that these are complex conjugates of each other leads to the equation
\begin{align}
	\left[4G^2-\left(t_1+\frac{u}{2}\,\rho\right)^2\right]\rho =0.
\end{align}
Solving for $\rho$, we arrive at the mean-field solution
\begin{equation}\label{eq:MFsol}
	\bar{\psi}_q=0,\quad
	\rho=\begin{cases}
	      0				& \!\!\text{for }\ t_1>2|G|,\\
	      \frac{2}{u}(2|G|-t_1)	& \!\!\text{for }\ t_1<2|G|.
	     \end{cases}
\end{equation}
To obtain $\bar{\psi}_c$, one simply substitutes $\rho$ into Eq.~\eqref{eq:solvingMF}.

Now, recalling that \cite{Kamenev2023,Sieberer2016-79}
\begin{align}\nonumber
	\bra\psi_{c,i}^*(t)\psi_{c,j}(t')\ket
	&\equiv \frac{1}{Z}\int\mathscr{D}[\psi_\pm^\pc,\psi_\pm^*]\, \psi_{c,i}^*(t)\psi^\pc_{c,j}(t')\, e^{-S}\\
	&= \Tr\big( \{\ha_i^\pdag(t),\ha_j^\dag(t')\}\,\dm_\text{ss}\big),
\end{align}
where $\dm_\text{ss}$ denotes the steady state of the system, the onset of the (space and time translation invariant) mean-field value \eqref{eq:MFsol} of $\rho=|\bar{\psi}_c|^2$ signals a macroscopic occupation of the zero-momentum mode.

So, the pumping and/or two-particle driving induce a transition to a Bose condensate phase at $t_1=2|G|$. In fact, this point also marks the boundary between the stable and unstable regions of the non-interacting part of the model ($u=0$); see Appx.~\ref{appx:stability} and Ref.~\cite{Barthel2021_12}. The two-particle loss ($u>0$) in the interacting model will stabilize the condensate.

\section{Tree-level scaling analysis}\label{sec:scaling}
The mean-field theory only describes the system correctly if the quartic terms in the action are irrelevant, i.e., above the upper critical dimension $d_c$. Let us perform the tree-level RG analysis to determine the engineering dimensions (a.k.a.\ canonical scaling dimensions) of the fields and coupling parameters such that we can assess the relevance of the (quartic) interaction terms and deduce $d_c$.
To this purpose, one considers the quadratic part of the action, i.e., the first three lines of Eq.~\eqref{eq:action-cq} and examines how quantities scale under a lowering of the ultraviolet cutoff $\Lambda\mapsto\Lambda/b$ (initially set by the inverse lattice spacing $1/a$), a corresponding rescaling of space/momenta $k\mapsto b k$, and renormalization $\psi_{c/q}\mapsto b^{[\psi_{c/q}]}\psi_{c/q}$, $g_i\mapsto b^{[g_i]}g_i$ of the field variables and coupling parameters. The engineering dimensions $[\psi_c]$, $[\psi_q]$, and $\{[g_i]\}$ are determined such that the action and low-momentum features of all Green's functions are invariant under this tree-level RG transformation.

With the  transformation
\begin{equation}\label{eq:Fourier}
	\phi_{c/q}(\vk,\omega):=\int_{\vec{x},t} e^{\mri(\omega t-\vk\cdot\vx)}\psi_{c/q}(\vx,t),
\end{equation}
to Fourier components, the quadratic part of the action can be easily diagonalized. The resulting free Green's function is calculated in Appx.~\ref{appx:freeGF}, and the dispersion relation for single-particle excitations from the steady state is found to be
\begin{equation}\label{eq:dispersion}
	\omega(\vk)=-\mri(t_1+K k^2)\pm\mri\sqrt{4G^2-J^2 k^4}.
\end{equation}
The many-body spectrum of quasi-free systems is fully determined by the single-particle dispersion \cite{Prosen2010-43,Barthel2021_12} and the \emph{dissipative gap} is
\begin{equation}\label{eq:gap}
	\Delta:=-\sup_{\lambda\neq 0\in \operatorname{spect}(\L)}\Re\lambda
	       =-\sup_\vk\, \Im \omega(\vk).
\end{equation}
So, the non-interacting system becomes gapless at the transition point $t_1=2|G|$, where the dissipative gap closes at $k=0$ with dispersion $\omega\sim -\mri K k^2$ for the slowest decay modes. Working with scaling dimension 1 for momenta, we hence have
\begin{equation}
	[k]=1\quad\text{and}\quad[\omega]=2.
\end{equation}

Let us now perform the tree-level scaling analysis for the critical point $t_1=2|G|$.
Due to the invariance of the partition function \eqref{eq:Z}, we have $[S]=0$. Using $[\ud^d x]=-d$ and $[\ud t]=-2$, it follows that
\begin{equation}\label{eq:scale1}\textstyle
	0=\big[\int_{\vx,t} \psi_c^*\partial_t\psi_q^\pc\big]=-d-2+[\psi_c]+2+[\psi_q].
\end{equation}
From this and the requirement that the terms with coefficients $t_1$, $K_1$, $K_2$, $J$, and $G$ in the action \eqref{eq:action-cq} are also scale invariant, it follows immediately that
\begin{equation}\label{eq:scale2}
	[t_1]=[G]=2\quad\text{and}\quad [K_1]=[J]=0.
\end{equation}

Now, on physical grounds it can be argued that $t_2$ should not scale, i.e., that we have a constant noise vertex in the action \cite{Sieberer2016-79} with
\begin{equation}\label{eq:scale3}\textstyle
	[t_2]=0\ \ \Rightarrow\ \
	0=\big[\int_{\vx,t} t_2\psi_q^*\psi_q^\pc\big]=-d-2+2[\psi_q].
\end{equation}
From this, Eq.~\eqref{eq:scale1}, and $0=\big[\int_{\vx,t} K_2 \psi_q^*\nabla^2\psi_q^\pc\big]$ we, finally conclude that
\begin{equation}\label{eq:dim}
	[\psi_c]=\frac{d-2}{2},\quad
	[\psi_q]=\frac{d+2}{2},\quad\text{and}\quad
	[K_2]=-2.
\end{equation}

As a consistency check, let us assess the free Green's function at the mean-field transition point as given in Eq.~\eqref{eq:freeGFcrit} of Appx.~\ref{appx:freeGF}. With the canonical dimensions of the coupling constants determined so far, we have the long-range and slow-frequency asymptotic behavior
\begin{align}
	\bra\phi_c^\pc\phi_c^* \ket\sim \frac{t_2}{K_1^2}\, k^{-d-6}  \ \ \text{and}\ \
	\bra\phi_c^\pc\phi_q^* \ket\sim \frac{1}  {K_1}\,   k^{-d-4},
\end{align} 
corresponding to
\begin{alignat*}{3}
	2[\phi_c]&=[t_2]-2[K_1]-d-6  \,\,&=-d-6, \\
	[\phi_c]+[\phi_q]&=-[K_1]-d-4&=-d-4.
\end{alignat*} 
Using the inverse of Eq.~\eqref{eq:Fourier} to transform back to real space and time, we have
\begin{align}
	2[\psi_c]=d-2\quad\text{and}\quad
	[\psi_c]+[\psi_q]=d,
\end{align}
which is indeed consistent with the previous result \eqref{eq:dim}.

With the canonical dimensions \eqref{eq:dim} of the fields and the condition $[S]=0$, we can determine the RG relevance of all terms in the action \eqref{eq:action-cq} of the interacting model. According to Eqs.~\eqref{eq:scale2}-\eqref{eq:dim}, the $t_1$ and $G$ terms are RG relevant, the $t_2,K_1$, and $J$ are marginal, and the $K_2$ term is irrelevant. For the quartic (interaction) terms, Eq.~\eqref{eq:dim} implies
\begin{align}\nonumber
	&\textstyle
	\big[\int_{\vx,t}\psi_c^2\psi_c^*\psi_q^*\big]=d-4,\quad
	\big[\int_{\vx,t}\psi_q^2\psi_c^*\psi_q^*\big]=d,\\
	&\textstyle\ \text{and}\quad
	\big[\int_{\vx,t}\psi_c\psi_c^*\psi_q\psi_q^*\big]=d-2.
\end{align}
Although these terms have the same coefficient $u$ in the bare action, they scale differently under RG due to the different scaling dimensions of $\psi_c$ and $\psi_q$. The upper critical dimension $d_c$ is defined such that all quartic terms are irrelevant for $d>d_c$. So, in this model, we have
\begin{equation}
	d_c=4,
\end{equation}
and the Gaussian fixed point
\begin{equation}\label{eq:GaussianFP}
	\tilde{t}_1=2|G|,\quad \tilde{u}=0
\end{equation}
with dynamical critical exponent $z=2$ [Eq.~\eqref{eq:dispersion}] is stable for $d>d_c$.

\section{Gaussian approximation above the upper critical dimension}\label{sec:Gaussian}
For $d>d_c$, let us now consider the Gaussian fluctuations around the mean-field solution \eqref{eq:MFsol}. Specifically, we substitute
\begin{equation}
	\psi_c=\bar{\psi}_c+\delta \psi_c\quad\text{and}\quad
	\psi_q=\delta \psi_q,
\end{equation}
and expand the action \eqref{eq:action-cq} to second order in the fluctuations $\delta \psi_c,\,\delta \psi_q$. The resulting Green's function is computed in Appx.~\ref{appx:Gaussian}, and we find the single-particle dispersion relation
\begin{equation}
	\omega_\pm=-\mri(t_1\!+K k^2\!+u\rho)\pm \mri\sqrt{\big|2\mri G-u\bar{\psi}_c^2/2 \big|^2\!\!-J^2k^4}.
\end{equation}
For $k=0$, one has $\omega_\pm=-\mri(t_1+u\rho \mp |2\mri G-u\bar{\psi}_c^2/2 |)$. So, the symmetric phase with $t_1>2|G|$ and $\rho=|\bar{\psi}_c|^2=0$ is gapped. For the symmetry-broken Bose condensate phase with $t_1<2|G|$ and $\rho>0$, let us first consider the case $G=0$, i.e., systems without two-particle driving. Then, the condensate has a gapless excitation with $\omega_+=0$ at $k=0$, corresponding to the Goldstone mode arising due to the spontaneous breaking of the continuous $\groupU(1)$ symmetry \eqref{eq:U1symmetry}. In contrast, with two-particle driving $G\neq 0$, both $\omega_+$ and $\omega_-$ are non-zero at $k=0$, meaning that this $\mc{PT}$-symmetry-broken Bose condensate is gapped. In summary, the symmetry broken phase is gapped for $G\neq0$ but gapless for $G=0$, and the symmetric phase is always gapped. 

There is a connection of these properties to general limitations on driven-dissipative phase transitions \Footnote{In analogy to phase transitions in closed systems \cite{Sondhi1997-69,Vojta2003-66,Sachdev2011}, we associate driven-dissipative phase transitions with a non-analytic change of the steady-state density operator \cite{Minganti2018-98}. According to general results on operator perturbation theory \cite{Kato1995}, this requires that the spectral gap $\Delta$ [Eq.~\eqref{eq:gap}] to the first excitation of the Liouvillian $\L$ closes at the transition point \cite{Kessler2012-86,Minganti2018-98}. Furthermore, we consider two gapped Liouvillians to be part of the same phase if one can construct a continuous path of gapped Liouvillians that links the two \cite{Zhang2022-129}. Note that this characterization does \emph{not} cover the (dynamical) dissipative topological phase transitions considered in Refs.~\cite{Altland2021-11,Bardyn2018-8,Caspel2019-6}, which are of a different nature.}
in quasi-free and quadratic open systems \footnote{We call a Markovian quantum system \emph{quasi-free} if the Hamiltonian is quadratic in fermionic or bosonic ladder operators and all Lindblad operators are linear; in addition, \emph{quadratic} open systems can have bilinear Hermitian Lindblad operators \cite{Barthel2021_12}.}: In their stable region, the steady states of such systems cannot undergo transitions between gapped phases unless one imposes symmetry constraints beyond invariance under single-particle basis transformations \cite{Zhang2022-129}. In fact, one can connect any two gapped quadratic Liouvillians through a continuous path of gapped Liouvillians by tuning single-particle loss terms.

The model that describes the order-parameter fluctuations around the mean-field solution for $d>d_c$ is Gaussian, i.e., quasi-free. As the fluctuations $\delta \psi_{c/q}$ are just linear in the (original) microscopic field variables $\psi_\pm$, we can also establish a direct connection between the gap-opening terms in the Gaussian model and corresponding terms in the quasi-free part of the original Liouvillian $\L$.
So, a phase transition between gapped phases could only occur inside or at the boundary to the unstable region of the non-interacting theory ($u=0$) and, indeed, the Bose condensation transition was found to occur at that boundary.

\section{RG analysis below the upper critical dimension}\label{sec:below}
Just below the upper critical dimension, the quartic term
\begin{equation}\label{eq:relevantInteract}
	\int_{\vec{x},t}\frac{u}{2}\left(\psi_c^2\psi_c^*\psi_q^*-\text{c.c.}\right)
\end{equation}
becomes relevant and can alter the phase diagram. We analyze this using an $\epsilon$ expansion \cite{Wilson1972-28,Wilson1972-28b,Wegner1973-8,Wilson1974-12,Altland2010} to perform the one-loop RG in $d=4-\epsilon$ dimensions. For simplicity, we set the two-particle driving $G=0$, which restores the $\groupU(1)$ symmetry \eqref{eq:U1symmetry}. 
We can drop all terms that have been identified as RG irrelevant in the tree-level scaling analysis, i.e., we only retain the interaction term \eqref{eq:relevantInteract} and also discard the $K_2$ term. For brevity, we use $K\equiv K_1$ in the following. 
Splitting the resulting action $S_R$ of all non-irrelevant terms into its Gaussian and quartic parts, we have
\begin{align}
	S_R&=S_G+S_u \quad \text{with} \\ \nonumber
	S_G&=\int_{\vx,t} \Big[ \psi_c^*\partial_t\psi_q^\pc+\psi_q^*\partial_t\psi_c^\pc
	  +(t_1\psi_q^*\psi_c^\pc-t_1^*\psi_c^*\psi_q^\pc)\\ \nonumber
	  &\!\!\!+(K-\mri J)\psi_c^*\nabla^2\psi_q^\pc
	   -(K+\mri J)\psi_q^*\nabla^2\psi_c^\pc
	   + t_2\psi_q^*\psi_q^\pc  \Big] ,\\ \nonumber
	S_u&=\frac{1}{2} \int_{\vx,t} \left[u\psi_c^2\psi_c^*\psi_q^*-u^*\psi_c^{*2}\psi_c^\pc\psi_q^\pc\right].
\end{align}
In the RG process, two additional terms are generated. The first, corresponds to a detuning term $\sim \ha^\dag\ha$ and the second to a Bose-Hubbard interaction term $\sim \ha^{\dag 2}\ha^2$ in the Hamiltonian. We have included them in the action $S_G$ and $S_u$, with coupling coefficients $i\Im t_1$ and $i\Im u$, respectively. So, while $t_1$ and $u$ are real in the initial model \eqref{eq:action-cq}, they generally flow to complex values during the RG.

As deduced in Appx.~\ref{appx:RG}, the one-loop RG flow equations for $S_R$ read
\begin{subequations}\label{eq:RG_flow}
\begin{align}
	\frac{\ud t_1}{\ud\ell} &=2t_1+\frac{\mc{S}_d t_2}{2 K+t_1+t_1^*}\, u + \mc{O}(u^2),\\
	\frac{\ud u}  {\ud\ell} &=\epsilon u-\frac{\mc{S}_d t_2}{2K^2}
	  \left(\!u^2\,\frac{3K+2\mri J}{2(K+\mri J)}+{uu^*}\!\right)+\mc{O}(u^3),
\end{align}
\end{subequations}
where we consider an infinitesimal momentum rescaling $k\to bk$ with $b=1+\ud\ell$, $\mc{S}_d:=2/[(4\pi)^{d/2}\Gamma(d/2)]$ is a phase-space factor, and we have set the ultraviolet cutoff to $\Lambda=1/a=1$. Here, we see that the edge-correlated pumping $\sim \gamma_p$ ($K$) is needed to make the theory renormalizable. The field renormalization has been chosen such that rates $t_2$ and $K$, as well as the inverse mass $J$ are RG invariant.
\begin{figure}[t]
	\includegraphics[width=0.96\columnwidth]{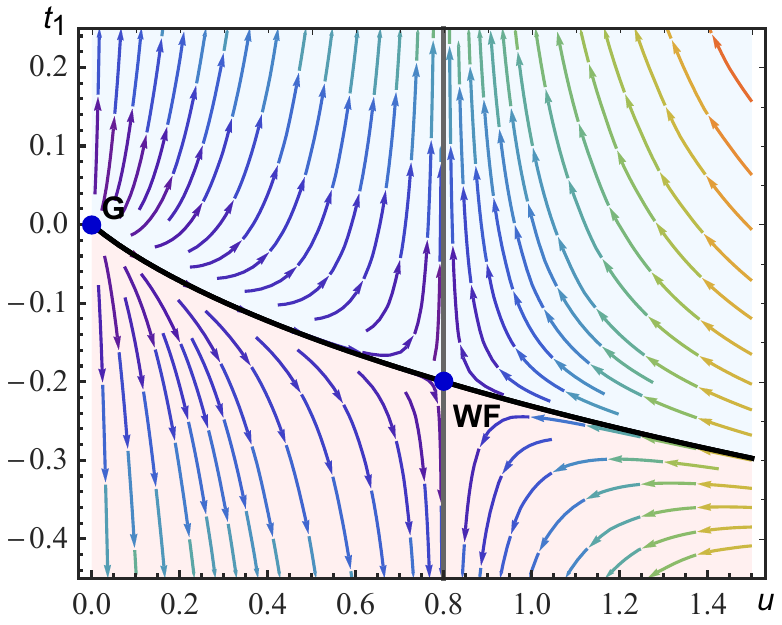}
	\caption{\label{fig:RG_flow}\textbf{RG flow diagram for $d<d_c$ and $J=G=0$.} The figure shows the RG flow \eqref{eq:RG_flow} for $d=3$ spatial dimensions which, for $J=0$, remains in the two-dimensional plane spanned by real $u$ and $t_1$. In addition to the unstable Gaussian fixed point (G), there is now a Wilson-Fisher-like fixed point (WF) which determines the critical behavior. The two-particle loss rate always flows to $\tilde{u}$ [Eq.~\eqref{eq:WilsonFisherFP}], while the difference of the single-particle loss and pumping $t_1=d(\gamma_l-\gamma_p)$ flows to plus or minus infinity depending on initial values. The critical manifold separates the symmetric phase (upper blue region) from the lower region with a finite steady-state condensate density. For the figure, we have chosen $K=\mc{S}_d t_2=1$.}
\end{figure}

The Gaussian fixed point at $t_1=u=0$ is stable for $d>d_c$ and the critical physics is described by the Gaussian field theory with dynamical critical exponent $z=2$ and the correlation-length exponent $\nu=1/2$ assuming their mean-field values.
For $d<d_c$, the Gaussian fixed point is unstable and the system now features an additional Wilson-Fisher-like fixed point at
\begin{subequations}\label{eq:WilsonFisherFP}
\begin{align}
	\tilde{t}_1&=-\epsilon\,\frac{K+\mri J}{5}+\mc{O}(\epsilon^2),
	\label{eq:WilsonFisherFP-t1}\\
	\tilde{u}&=\epsilon\,\frac{4K(K+\mri J)}{5\mc{S}_d t_2}+\mc{O}(\epsilon^2).
	\label{eq:WilsonFisherFP-u}
\end{align}
\end{subequations}

To analyze the flow in the vicinity of this point, we express $t_1=\tilde{t}_1+\delta t_1$ and $u=\tilde{u}+\delta u$ and expand the flow equations \eqref{eq:RG_flow} to linear order in the deviations from the fixed point \eqref{eq:WilsonFisherFP}, finding
\begin{equation*}
	\frac{\ud}{\ud \ell}
	\Pmatrix{\Re\delta t_1 \\ \Im\delta t_1  \\  \Re\delta u   \\   \Im\delta u}
	=
	\begingroup\renewcommand\arraystretch{1.3}
	\Bmatrix{
	2-\frac{2\epsilon}{5}     & 0 & * & * \\
	-\,\frac{2\epsilon J}{5K} & 2 & * & * \\
	0 & 0 & -\epsilon & 0 \\
	0 & 0 & -\,\frac{4J\epsilon}{5K}  &-\,\frac{\epsilon}{5}}
	\endgroup
	\Pmatrix{\Re\delta t_1 \\ \Im\delta t_1  \\  \Re\delta u   \\   \Im\delta u}.
\end{equation*}
The upper-right block is $\mc{S}_d t_2\id_{2\times 2}/2K+\mc{O}(\epsilon)$ and does not affect the eigenvalues of the matrix. The flow of $\delta u$ is independent of $\delta t_1$ and is characterized by the eigenvalues
\begin{equation}
	\lambda_3=-\epsilon+\mc{O}(\epsilon^2)\quad\text{and}\quad
	\lambda_4=-\epsilon/5+\mc{O}(\epsilon^2)
\end{equation}
of the lower-right $2\times 2$ submatrix. So, $u$ will always flow towards the fixed-point value \eqref{eq:WilsonFisherFP-u}. Since the generating matrix of the linearized RG flow is already in block-triangular form, the remaining two eigenvalues can be read off directly as 
\begin{align}\label{eq:t1-eigenvals}
	\lambda_1=2-2\epsilon/5+\mc{O}(\epsilon^2)\quad\text{and}\quad\lambda_2=2+\mc{O}(\epsilon^2).
\end{align}
These correspond to two relevant directions concerning the real and imaginary parts of $t_1$. However, as already pointed out in Ref.~\cite{Sieberer2016-79}, the $\groupU(1)$ symmetry \eqref{eq:U1symmetry} of the model for $G=0$ can be used to impose a gauge where $t_1$ is real by going to a suitable rotating frame. In particular, the transformation
\begin{equation}
	\phi_{c/q}(\vx,t) \mapsto \phi_{c/q}(\vx,t) e^{-\mri \omega_0 t}
\end{equation}
generates the term $-\mri(\omega_0\psi_c^*\psi_q^\pc+\psi_q^*\psi_c^\pc)$ in $S_G$, which cancels the term $\propto\Im t_1$ for $\omega_0=\Im t_1$.
With real $t_1$, we are left with only one physically relevant direction, determining the boundary between the normal state and the condensate in the remaining three-dimensional parameter space. The corresponding eigenvalue $\lambda_1$ in Eq.~\eqref{eq:t1-eigenvals} yields the correlation-length exponent \cite{Taeuber2014-4}
\begin{equation}
	\nu=\frac{1}{\lambda_1}=\frac{1}{2}+\frac{\epsilon}{10}+\mc{O}(\epsilon^2)
	\ \ \text{such that}\ \
	\xi\sim|\delta t_1|^{-\nu}
\end{equation}
for the correlation length near the critical point.
To see this, note that, for $u=\tilde{u}$ and a rescaling factor $b=e^\ell$, the RG flow equations imply that two-point correlation functions obey homogeneity relations of the form \cite{Sachdev2011}
\begin{equation}
	C(e^{-\ell}\Delta\vx,e^{\lambda_1\ell}\delta t_1)=e^{(d-2+\eta)\ell}C(\Delta\vx,\delta t_1).
\end{equation}
Evaluating this with $\ell$ chosen such that $e^{\lambda_1\ell}\delta t_1=\pm 1$ gives the scaling form
\begin{equation*}
	C(\Delta\vx,\delta t_1)=\xi^{-(d-2+\eta)}F_\pm\Big(\frac{\Delta\vx}{\xi}\Big)
	\ \ \text{with}\ \
	\xi=|\delta t_1|^{-\frac{1}{\lambda_1}}.
\end{equation*}

The RG flow is illustrated in Fig.~\ref{fig:RG_flow}. Its structure is very similar to that of the celebrated Wilson-Fisher phase diagram \cite{Wilson1972-28,Wilson1974-12,Altland2010} in the scalar $\phi^4$ theory \footnote{The scalar $\phi^4$ theory, is the only interacting scalar field theory of physical interest that is also renormalizable. It describes, for example, the $d$-dimensional Ising model close to its critical point and, more generally, the critical behavior of classical systems with a single order parameter (cf.\ Ginzburg-Landau theory of second-order phase transitions). Higher-order terms or terms involving derivatives turn out to be RG irrelevant.}. Depending on its initial value, $t_1$ will flow to $+\infty$ or $-\infty$, separating the symmetric and Bose condensate phases. The critical manifold lies in the unstable region of the non-interacting theory (cf.\ Appx.~\ref{appx:stability}).

\section{Discussion}\label{sec:discussion}
We have seen how incoherent pumping and/or coherent two-particle driving in competition with single-particle and two-particle loss can stabilize a Bose-Einstein condensate as a non-equilibrium steady state. Above the upper critical dimension $d_c=4$ of the associated driven-dissipative phase transition, the fluctuations around the mean-field solution are captured by a Gaussian theory. According to a general result discussed in Ref.~\cite{Zhang2022-129}, transitions between two gapped phases can never occur inside the stable region of a non-interacting Markovian theory. For $d>d_c$, our bosonic system is a specific example. The condensation transition then occurs right at the boundary between the stable and unstable regions of the non-interacting theory. With two-particle driving, the condensate is gapped, i.e., we have a transition between two distinct gapped phases. Without two-particle driving, the $\groupU(1)$ symmetry results in gapless Goldstone-mode excitations from the steady-state condensate.

For systems below the upper critical dimension ($d<d_c$), the Gaussian fixed point becomes unstable, and we have carried out the one-loop RG analysis using $\epsilon$ expansion. Interestingly, the transition still occurs in the unstable region of the non-interacting theory, and the physics of the critical point is described by the universal field theory of a Wilson-Fisher-like fixed point \cite{Wilson1972-28,Wilson1974-12,Altland2010,Taeuber2014-4}.
As shown in Sec.~\ref{sec:below}, coupling coefficients in the Keldysh action can flow to complex values during the RG. This is due to the non-Hermiticity of the Liouvillian, which is an intrinsic property of open systems. The one-loop analysis yields a correlation-length exponent of $\nu=1/2+(d_c-d)/10+\mc{O}\left((d_c-d)^2\right)$. The value $\nu=0.6$ for $d=3$ dimensions lies between the mean-field value $1/2$ and the value $\nu_\text{fRG}\approx 0.716$ found in the functional-RG analysis by Sieberer \emph{et al.}\ \cite{Sieberer2013-110,Sieberer2014-89}.

Let us shortly contrast these results with BEC in closed systems, where we are dealing with a single complex field $\psi$: As honored with the 2001 Nobel Prize, a dilute interacting Bose gas can undergo BEC at low temperatures, where a single-particle state gets macroscopically occupied \cite{Anderson1995-269,Davis1995-75}. The transition to the normal (symmetric) phase is caused by thermal fluctuations. The long-range physics of nonzero-temperature transitions in $d$-dimensional quantum systems are described by \emph{classical} field theories in $d$ dimensions \cite{Sondhi1997-69,Vojta2003-66,Sachdev2011} \footnote{In the path-integral framework, a $d$-dimensional quantum system is mapped to a $(d+1)$-dimensional classical system with an additional periodic imaginary-time direction of length $\beta=1/k_B T$. Correlation lengths $\xi$ and times $\xi_\tau\sim\xi^z$ diverge when approaching the critical temperature such that relevant field configurations are basically locked in the $\tau$ direction when $\xi_\tau\gg \beta$ and the long-range physics is then  described by a classical field theory in $d$ dimensions.} In the case of BEC with the $\groupU(1)$ symmetry \eqref{eq:U1symmetry}, this is the $\groupO(2)$ model (a.k.a.\ $XY$ model) which has upper critical dimension $d_c=4$ and correlation-length exponent $\nu\approx 0.67$ for $d=3$ as determined theoretically \cite{Gorishny1984-101,Guida1998-31,Jasch2001-42,Campostrini2006-74} and in experiments \cite{Lipa2003-68,Donner2007-315}.
Condensation can also be driven by quantum fluctuations in closed systems at zero temperature.
Such quantum phase transitions can be captured by classical $(d+1)$-dimensional field theories \cite{Sondhi1997-69,Vojta2003-66,Sachdev2011}.
In the case of the Bose-Hubbard model, the competition between the coherent kinetic and on-site repulsion terms leads to a transition between the Mott insulator and a superfluid (BEC) \cite{Fisher1989-40,Sachdev2011}. Coming from a Mott lobe with integer particle density $\rho$, one has to distinguish two cases. Generic transitions with continuously changing $\rho$ are in the dilute-Bose-gas universality class with dynamical exponent $z=2$ (quadratic dispersion for the excess particles), $d_c=2$, and the mean-field value $1/2$ for $\nu$ in $d\geq 2$ dimensions. Transitions with fixed $\rho$ are described by the classical $(d+1)$-dimensional $\groupO(2)$ model with $z=1$ due to space-time isotropy (linear dispersion), $d_c+1=4$, and the mean-field value $\nu=1/2$ in $d\geq 3$ dimensions \cite{Fisher1989-40,Sachdev2011}.

It would be valuable to probe the field-theoretical predictions for the driven-dissipative BEC in numerical simulations. To this purpose it may be useful to consider the limit $u\to\infty$ of infinitely strong two-particle loss, restricting the maximum number of bosons per site to one. Above the upper critical dimension, the $u$ term is RG irrelevant and should not affect the critical behavior.

\begin{acknowledgments}
We gratefully acknowledge valuable discussions with Sebastian Diehl, Mikhail Pletyukhov, and Toma\v{z} Prosen.
\end{acknowledgments}

\appendix

\onecolumngrid
\section{Experimental realization}\label{appx:experiment}
The model \eqref{eq:model} can, for example, be realized in circuit QED systems \cite{Blais2004-69,Wallraff2004-431,Hartmann2006-2,Carusotto2013-85,Hartmann2016-18,Fitzpatrick2017-7,Blais2021-93}, consisting of a lattice of interconnected microwave resonators. Each resonator contributes a localized harmonic photon mode associated with bosonic annihilation and creation operators $\ha_i$ and $\ha_i^\dag$. Evanescent coupling between neighboring cavities due to the spatial overlap of their modes gives rise to the photon-hopping term in the Hamiltonian \cite{Greentree2006-2,Fitzpatrick2017-7}. Nondissipative photon interactions can be generated through Kerr-type nonlinearities by embedding a superconducting qubit (an artificial two or multi-level atom) in each cavity \cite{Blais2004-69,Wallraff2004-431,Blais2021-93}. To not complicate the model unnecessarily, we have not included such $\ha_i^\dag\ha_i^\dag\ha_i\ha_i$ Hamiltonian terms. However, they generally arise during the RG flow ($u$ attains an imaginary part).
Two-photon driving, yielding the term $\ha^2+\ha^{\dag 2}$ in the Hamiltonian after going to the rotating frame, can be realized in different ways \cite{Yamamoto2008-93,Leghtas2015-347,Zhao2018-10}, e.g., by terminating a $\lambda/4$ microwave resonator with a flux-pumped SQUID (a Josephson parametric amplifier). Generally, this involves an external microwave drive at twice the cavity frequency.

Let us now turn to the dissipative components of the model. In typical quantum-optical and circuit-QED experiments, the system is weakly coupled to the (electromagnetic) environmental degrees of freedom, and we have a clear separation between characteristic system time scales and the scale for the decay of time-time correlation functions in the environment \cite{Carmichael1993,Gardiner2004,Blais2021-93}. Hence, the Born-Markov-secular approximation \cite{Bloch1957-105,Redfield1957-1,Redfield1965-1,Davies1974-39,Davies1976-219,Duemcke1979-34,Walls1994,Breuer2007} is applicable and allows us to describe the system with a Markovian Lindblad master equation \eqref{eq:Lindblad} for the system density operator \cite{Lindblad1976-48,Gorini1976-17}.

One-photon losses are due to the finite quality factor of the resonators and result in the dissipators with Lindblad operators $\ha_i$. Two-photon losses ($\ha_i^2$) naturally arise both due to Kerr-type nonlinearities and the engineered two-photon driving \cite{Leghtas2015-347}. In our study, the role of the two-particle loss, however small, is to stabilize the condensate. Incoherent single-photon pumping leads to the Lindblad operators $\ha_i^\dag$. The edge-correlated pumping with Lindblad operators $\ha_i^\dag+\ha_j^\dag$ on nearest-neighbor sites can, for example, be implemented either by a set of incoherently pumped auxiliary cavities or superconducting qubits, each coupled to one pair of neighboring resonators, as suggested in Refs.~\cite{Porras2019-122} and \cite{Marino2016-116}, respectively. Operating the auxiliary cavities or qubits in a fast-decaying regime, their dynamics can be adiabatically eliminated \cite{Cirac1992-46,Breuer2007}.

\section{The free bosonic Green's function}\label{appx:freeGF}
To determine the free Green's function of the Keldysh action Eq.~\eqref{eq:action-cq}, i.e., the Green's function of its quadratic part, we can follow the steps outlined in Ref.~\cite{Sieberer2016-79}. With $k:=|\vk|$, the quadratic part of the action reads
\begin{align}\nonumber
	S_G&=
	\int_{\vx,t} \Big[ \psi_c^*\partial_t\psi_q^\pc+\psi_q^*\partial_t\psi_c^\pc
	  -t_1(\psi_c^*\psi_q^\pc-\psi_q^*\psi_c^\pc)
	  + t_2\psi_q^*\psi_q^\pc
	  +K_1\left(\psi_c^*\nabla^2\psi_q^\pc -\psi_q^*\nabla^2\psi_c^\pc \right ) +2K_2\psi_q^*\nabla^2\psi_q^\pc \\ \nonumber
	&\qquad\qquad
	  -\mri J\left( \psi_c^*\nabla^2\psi_q^\pc+\psi_q^*\nabla^2\psi_c^\pc \right)
	  +2\mri G(\psi_c\psi_q+\psi_c^*\psi_q^*)\Big]  \nonumber \\
	&\stackrel{\eqref{eq:Fourier}}{=}
	 \int_{\vk,\omega>0}\Big[\left(-\mri\omega+\mri J k^2\right)\big(\phi_c^*(\vk,\omega)\phi_q^\pc(\vk,\omega)+c.c.\big)
	     -(t_1+K_1 k^2)\big(\phi_c^*(\vk,\omega)\phi_q^\pc(\vk,\omega) - c.c.\big) \nonumber \\
	&\qquad\qquad\quad
	  +(t_2-2K_2 k^2)\phi_q^*(\vk,\omega)\phi_q^\pc(\vk,\omega) +2\mri G\, \big(\phi_c(\vk,\omega)\phi_q(-\vk,-\omega)+\phi_q^*(\vk,\omega)\phi_c^*(-\vk,-\omega)\big)
	  \Big].
\end{align}
Since the super-particle number is not conserved when $G\neq 0$, we define the Nambu-Keldysh spinor
\begin{equation}
	\vphi(\vk,\omega)
	:= \Pmatrix{\phi_c(\vk,\omega),\, \phi_c^*(-\vk,-\omega),\,
	            \phi_q(\vk,\omega),\, \phi_q^*(-\vk,-\omega) }^\intercal
\end{equation}
such that the Gaussian action can be written as
\begin{align}
	S_G &= \int_{\vk,\omega>0} \vphi^\dag (\vk,\omega) P(\vk,\omega) \vphi (\vk,\omega)
\end{align}
with the inverse Green's function
\begin{align}
	P(\vk,\omega) 
	&=\Bmatrix{0&P_A(\vk,\omega)\\P_R(\vk,\omega)&P_K(\vk,\omega)}
	 =\Bsmatrix{
	  0              & 0                &-\mri\omega+\mri J k^2-t_1-K_1 k^2   &2\mri G \\
	  0              & 0                &2\mri G  &\mri\omega+\mri J k^2+t_1+K_1 k^2 \\
	  -\mri\omega+\mri J k^2+t_1+K_1 k^2&2\mri G  & t_2-2K_2 k^2 & 0 \\
	  2\mri G & \mri\omega+\mri J k^2-t_1-K_1 k^2 & 0            & t_2-2K_2 k^2 },
\end{align}
where $P_K^\dag(\vk,\omega)=P_K(\vk,\omega)$ and $P_R(\vk,\omega)=-P_A^\dag(\vk,\omega)$.

Now, using the general property of Gaussian integrals,
\begin{align}
	Z[\vb,\vb^\dag]
	&:= \frac{1}{Z}\int \mathscr{D}[\psi,\psi^\dag]\, e^{-\vpsi^\dag A \vpsi+\vb^\dag\vpsi+\vpsi^\dag \vb}
	  =e^{\vb^\dag A^{-1}  \vb}\quad\text{such that}  \nonumber \\
	\bra \vpsi\vpsi^\dag \ket
	 &=\frac{1}{Z}\int \mathscr{D}[\psi,\psi^\dag]\, \vpsi\vpsi^\dag e^{-\vpsi^\dag A \vpsi}
	  =\left.  \frac{\partial^2 Z[\vb,\vb^\dag]}{\partial \vb^\dag \partial \vb }\right|_{\vb=\vb^\dag=0}=A^{-1},
	\label{eq:gaussian}
\end{align}
we have
\begin{align}
	&\left\langle\vphi(\vk,\omega)\vphi^\dag(\vk',\omega') \right\rangle=\delta_{\vk,\vk'}\delta_{\omega,\omega'} P^{-1}(\vk,\omega)\equiv \delta_{\vk,\vk'}\delta_{\omega,\omega'} G(\vk,\omega).
\end{align}
When going to the thermodynamic limit and considering infinite evolution times, $\delta_{\vk,\vk'}$ and $\delta_{\omega,\omega'}$ are given by Dirac delta distributions,
\begin{align}
	\delta_{\vk,\vk'}=(2\pi)^d\delta^d(\vk-\vk'),\quad
	\delta_{\omega, \omega'}=2\pi \delta(\omega-\omega').
\end{align} 
This gives the free Green's function 
\begin{align}
	\bra\vphi(\vk,\omega)\vphi^\dag(\vk',\omega')\ket 
	&= \Bsmatrix{
	\bra\phi_c(k_\mu) \phi_c^*(k'_\mu) \ket&
	\bra\phi_c(k_\mu) \phi_c (-k'_\mu) \ket&
	\bra\phi_c(k_\mu) \phi_q^*(k_\mu') \ket&
	\bra\phi_c(k_\mu) \phi_q (-k'_\mu) \ket\\
	\bra\phi_c^*(-k_\mu) \phi_c^*(k'_\mu) \ket&
	\bra\phi_c^*(-k_\mu) \phi_c (-k'_\mu) \ket&
	\bra\phi_c^*(-k_\mu) \phi_q^*(k_\mu') \ket&
	\bra\phi_c^*(-k_\mu) \phi_q (-k'_\mu) \ket\\
	\bra\phi_q(k_\mu) \phi_c^*(k'_\mu) \ket&
	\bra\phi_q(k_\mu) \phi_c (-k'_\mu) \ket&
	\bra\phi_q(k_\mu) \phi_q^*(k_\mu') \ket&
	\bra\phi_q(k_\mu) \phi_q (-k'_\mu) \ket\\
	\bra\phi_q^*(-k_\mu) \phi_c^*(k'_\mu) \ket&
	\bra\phi_q^*(-k_\mu) \phi_c (-k'_\mu) \ket&
	\bra\phi_q^*(-k_\mu) \phi_q^*(k_\mu') \ket&
	\bra\phi_q^*(-k_\mu) \phi_q (-k'_\mu) \ket }\nonumber\\
	&=\delta_{\vk,\vk'}\delta_{\omega,\omega'}
	\Bmatrix{ G_K(\vk,\omega) & G_R(\vk,\omega)\\
	          G_A(\vk,\omega) & 0}
	\label{eq:freeGF}
\end{align}
with the Keldysh and retarded Green's functions
\begin{align}
G_K(\vk,\omega)&=\textstyle
	\frac{1}{\big|J^2 k^4+(t_1+K_1 k^2-\mri\omega)^2-4G^2\big|^2}
	\Bsmatrix{
	(t_2-2K_2 k^2)\left((\omega+J k^2)^2+(t_1+K_1 k^2)^2+4G^2\right)&
	-4\mri G(t_2-2K_2 k^2)(t_1+K_1 k^2-\mri J k^2) \vspace{0.3em} \\
	 4\mri G(t_2-2K_2 k^2)(t_1+K_1 k^2+\mri J k^2) &
	(t_2-2K_2 k^2)\left((\omega-J k^2)^2+t_1^2+4G^2\right)}, \nonumber \\
G_R(\vk,\omega)&=\textstyle
	\frac{1}{J^2 k^4+(t_1+K_1 k^2-\mri\omega)^2-4G^2}
	\Bsmatrix{
	 -\mri \omega-\mri J k^2+t_1+K_1 k^2 & 2\mri G \vspace{0.3em} \\
	2\mri G &
	  \mri \omega-\mri J k^2-t_1-K_1 k^2}.
\end{align}
In accordance with the general causal structure of Keldysh actions \cite{Kamenev2023}, we have $G_K=G_K^\dag$ and $G_A(\vk,\omega)=-G_R^\dag(\vk,\omega)$ for the advanced Green's function.

The single-particle excitation spectrum of the Gaussian (non-interacting) system is determined by zeros in the determinant of the inverse retarded Green's function,
\begin{equation}\label{eq:freeDisp}
	\det P_R(\vk,\omega)=0\quad\Leftrightarrow\quad
	\omega(\vk)=-\mri(t_1+K_1 k^2)\pm\mri\sqrt{4G^2-J^2 k^4}.
\end{equation}
So, the dissipative gap closes at $k=0$  for $t_1=2|G|$, and the low-lying excitations from the steady state (the slowest decay modes) have the dispersion
\begin{equation}
	\omega(\vk)\stackrel{t_1=2|G|}{=}
	-\mri K_1 k^2-\frac{\mri J^2}{2 t_1}\,k^4+\mc{O}(k^6).
\end{equation}
At this point, the Green's functions are
\begin{align}
G_K(\vk,\omega)&= \textstyle
	\frac{1}{\big|J^2 k^4+(K_1 k^2-\mri\omega)^2+2t_1(K_1 k^2-\mri\omega)\big|^2}
	\Bsmatrix{
	(t_2-2K_2 k^2)\left((\omega+J k^2)^2+(t_1+K_1 k^2)^2+4G^2\right) &
	4\mri G(t_2-2K_2 k^2)(-t_1+\mri J k^2)	\vspace{0.3em} \\
	4\mri G(t_2-2K_2 k^2)( t_1+\mri J k^2) &
	(t_2-2K_2 k^2)\left((\omega-J k^2)^2+t_1^2+4G^2\right) }, \nonumber \\
G_R(\vk,\omega)&= \textstyle
	\frac{1}{J^2 k^4+(K_1 k^2-\mri\omega)^2+2t_1(K_1 k^2-\mri\omega)}
	\Bsmatrix{
	-\mri \omega-\mri J k^2+t_1+K_1 k^2 & 2\mri G \vspace{0.3em} \\
	2\mri G & \mri \omega-\mri J k^2-t_1-K_1 k^2 }
	\label{eq:freeGFcrit}
\end{align}
With this result, one can assess the scaling behavior and the upper-critical dimension as discussed in Sec.~\ref{sec:scaling}.

\section{Stability boundary of the non-interacting system}\label{appx:stability}
The system is stable as long as all excitations decay or oscillate with constant amplitude at large times. For the field-theoretical description of the non-interacting system discussed in Appendix~\ref{appx:freeGF}, stability implies that all single-particle frequencies $\omega(\vk)$ in Eq.~\eqref{eq:freeDisp} must have non-positive imaginary parts. At $k\equiv|\vk|=0$, The square root in Eq.~\eqref{eq:freeDisp} is $2|G|$ such that the system is unstable when $t_1<2|G|$.

We can also determine the range of stability for the quasi-free (quadratic) part $\L_G$ of the original lattice model \eqref{eq:model} with Liouvillian
\begin{align}\label{eq:LG}
	\L_G(\dm)&=-\mri[\hat{H} ,\dm] +2d\gamma_l \sum_i \D[\ha_i](\dm)  +\frac{\gamma_p}{2}\sum_{\bra i,j\ket} \D[\ha_i^\dag+\ha_j^\dag](\dm).
\end{align}
Employing the formalism described in Refs.~\cite{Barthel2021_12, Zhang2022-129}, the many-body spectrum of $\L_G$ is determined by the (single-particle) eigenvalues of the $2\times 2$ matrices
\begin{equation}%\textstyle
	\mri\tilde{x}(\vk)=2\sigma^y\tilde{h}(\vk)-\sigma^y\tilde{b}_\text{i}(\vk)
	 = \mri\Big[-d\gamma_l+ \frac{\gamma_p}{2}\sum_{j=1}^d\big(1+\cos k_j \big)\Big]\,\id +2\mri G\,\sigma^x
	    +2\tilde{J}\sum_{j=1}^d(1-\cos k_j)\,\sigma^y,
\end{equation}
where $\sum_j$ denotes the sum over all spatial dimensions and $\sigma^x,\sigma^y$ are the Pauli matrices. The eigenvalues of $\mri\tilde{x}(\vk)$ are
\begin{align}\label{eq:freeDispLattice}
	-\mri d\gamma_l+ \frac{\mri\gamma_p}{2}\sum_{a}\big(1+\cos k_j \big)
	\pm\mri \sqrt{\textstyle 4G^2-4\tilde{J}^2\left[\sum_j(1-\cos k_j)\right]^2}.
\end{align}
Their maximum imaginary part $-d(\gamma_l-\gamma_p)+2|G|=-t_1+2|G|$ is obtained at quasi-momentum $k=0$. The system is unstable if the maximum imaginary part is positive and stable otherwise \cite{Barthel2021_12}. Thus, we find the same stability boundary $t_1=2|G|$ as in the field-theoretical description. In fact, Eq.~\eqref{eq:freeDispLattice} agrees with Eq.~\eqref{eq:freeDisp} when substituting $k\mapsto a k$ to account for the lattice spacing, expanding the cosines to second order, and plugging in the definitions from Eq.~\eqref{eq:parameterDef}.

The transition point suggested by the mean-field analysis in Sec.~\ref{sec:MFT} coincides with this stability boundary of the non-interacting theory. Without two-particle driving ($G=0$), the boundary is at $t_1=0$. So, the two-particle driving term in the Hamiltonian can induce instability.

\section{Gaussian approximation}\label{appx:Gaussian}
Let us now turn to the Gaussian approximation. Substituting $\psi_c=\bar{\psi}_c+\delta \psi_c$ and $\psi_q=\delta \psi_q$ and expanding the action \eqref{eq:action-cq} up to second order in the fluctuations $\delta \psi_c,\delta \psi_q$ around the mean-field solution \eqref{eq:MFsol}, we obtain the effective quadratic action
\begin{align} \nonumber
	S_\eff=
	\int_{\vx,t}& \Big[ \delta\psi_c^*\partial_t \delta\psi_q^\pc+\delta\psi_q^*\partial_t \delta\psi_c^\pc
	  -t_1(\delta\psi_c^*\delta\psi_q^\pc-\delta\psi_q^*\delta\psi_c^\pc)
	  +t_2 \delta\psi_q^*\delta\psi_q^\pc
	  +K_1\left( \delta\psi_c^*\nabla^2\delta\psi_q^\pc -\delta\psi_q^*\nabla^2\delta\psi_c^\pc\right)\\  \nonumber
	& +2K_2\delta\psi_q^*\nabla^2\delta\psi_q^\pc
	  -\mri J\left(\delta\psi_c^*\nabla^2 \delta\psi_q^\pc+\delta\psi_q^*\nabla^2 \delta\psi_c^\pc \right)
	  +2\mri G(\delta\psi_c\delta\psi_q+\delta\psi_c^*\delta\psi_q^*) \\
 	& +\frac{u}{2}(\bar{\psi}_c^2\delta\psi_c^*\delta\psi_q^* +2\rho \delta\psi_c^\pc\delta\psi_q^*-\text{c.c.})
 	  +2u\rho\delta\psi_q^*\delta\psi_q^\pc \Big].
\end{align} 
We proceed as in Appx.~\ref{appx:freeGF}, transforming to momentum space,
\begin{align}
	S_\eff&\stackrel{\eqref{eq:Fourier}}{=}
	 \int_{\vk,\omega>0} \delta\vphi^\dag(\vk,\omega)P(\vk,\omega)\delta\vphi(\vk,\omega),
\end{align}
with the Nambu-Keldysh spinor
$\delta\vphi(\vk,\omega):= \left( \delta\phi_c(\vk,\omega), \delta\phi_c^*(-\vk,-\omega), \delta\phi_q(\vk,\omega), \delta\phi_q^*(-\vk,-\omega) \right)^T$ and the
kernel $P(\vk,\omega)$ given by the inverse Green's function
\begin{subequations}
\begin{align}
	P(\vk,\omega)&=
	\Bmatrix{
	0 & P_A(\vk,\omega)   \\
	P_R(\vk,\omega) & P_K(\vk,\omega) },
	\quad \text{where}\\
	P_R(\vk,\omega)&=
	\Bmatrix{
	-\mri\omega +\mri J k^2+t_1+K_1 k^2+u\rho  &2\mri G-\frac{u}{2}\bar{\psi}_c^2  \\
	2\mri G+\frac{u}{2}\bar{\psi}_c^{*2}  &\mri\omega +\mri J k^2-t_1-K_1 k^2-u\rho }
	= -P_A^\dag(\vk,\omega),
	\quad \text{and}\\
	P_K(\vk,\omega)&=
	\Bmatrix{
	t_2-2K_2 k^2+2u\rho  & 0\\
	0  & t_2 -2K_2 k^2+2u\rho } .
\end{align}
\end{subequations}
The single-particle excitation spectrum in the Gaussian approximation is determined by zeros in the determinant of the inverse retarded Green's function,
\begin{equation}
	\det P_R(\vk,\omega)=0\quad\Leftrightarrow\quad
	\omega(\vk)=-\mri(t_1+K_1 k^2+u\rho)\pm \mri\sqrt{|2\mri G-u\bar{\psi}_c^2/2 |^2-J^2k^4}.
\end{equation}

\section{One-loop renormalization group}\label{appx:RG}
In this section, we carry out the one-loop RG analysis for the interacting bosonic model \eqref{eq:action-cq} with $G=0$ and an $\epsilon$ expansion for $d=4-\epsilon$ spatial dimensions. As discussed at the beginning of Sec.~\ref{sec:below}, we  drop all terms that have been identified as RG irrelevant in the tree-level scaling analysis. Relabeling $K\equiv K_1$ (the $K_2$ term is RG irrelevant), our starting point is the action $S_R$ of all non-irrelevant terms
\begin{subequations}\label{eq:SR}
\begin{align}
	S_R&=S_G+S_u, \quad \text{with} \\ \label{eq:SR-G}
	S_G&=\int_{\vx,t} \Big[ \psi_c^*\partial_t\psi_q^\pc+\psi_q^*\partial_t\psi_c^\pc
	  +(t_1\psi_q^*\psi_c^\pc-t_1^*\psi_c^*\psi_q^\pc)
	  +(K-\mri J)\psi_c^*\nabla^2\psi_q^\pc
	  -(K+\mri J)\psi_q^*\nabla^2\psi_c^\pc
	  + t_2\psi_q^*\psi_q^\pc  \Big] ,\\
	S_u&=\frac{1}{2} \int_{\vx,t} \left[u\psi_c^2\psi_c^*\psi_q^*-u^*\psi_c^{*2}\psi_c^\pc\psi_q^\pc\right],
\end{align}
\end{subequations}
which we have split into its Gaussian and quartic parts. While $t_1$ and $u$ are real in the original model, we will find them to generally flow to complex values during the RG. With $G=0$, the $\groupU(1)$ symmetry is restored such that all expectation values $\bra\phi_{c/q}(k_\mu)\phi_{c/q}(-k_\mu')\ket$ are zero, and the free Green's function \eqref{eq:freeGF} simplifies to
\begin{align}
	\Bmatrix{
	   \bra\phi_c(\vk,\omega)\phi_c^*(\vk',\omega') \ket& \bra\phi_c(\vk,\omega)\phi_q^*(\vk',\omega')\ket \\
	   \bra\phi_q(\vk,\omega)\phi_c^*(\vk',\omega') \ket& \bra\phi_q(\vk,\omega)\phi_q^*(\vk',\omega') \ket}
	 =\delta_{\vk,\vk'}\delta_{\omega,\omega'}
	  \Bmatrix{
	    \frac{t_2}{|-\mri\omega+\mri J k^2+t_1+K k^2|^2} & \frac{1}{-\mri\omega+\mri J k^2+t_1+K k^2} \\
	    \frac{1}{-\mri\omega+\mri J k^2-t_1^*-K k^2} & 0}.
	\label{eq:gf-appx}
\end{align}

The first step of the RG analysis is to decompose the fields into slow and fast components,
\begin{align}
	\psi_{c/q<}(\vx,t)&:= \int_{0}^{\Lambda/b}\frac{\ud^dk}{(2\pi)^d}\int_{-\infty}^\infty\frac{\ud\omega}{2\pi}\,
	                        e^{\mri\vk\cdot\vx-\mri\omega t} \phi_{c/q}(\vk,\omega), \nonumber \\
	\psi_{c/q>}(\vx,t)&:= \int_{\Lambda/b}^{\Lambda}\frac{\ud^dk}{(2\pi)^d}\int_{-\infty}^\infty\frac{\ud\omega}{2\pi}\,
	                        e^{\mri\vk\cdot\vx-\mri\omega t} \phi_{c/q}(\vk,\omega),
\end{align}
where, $b\gtrsim 1$ is the rescaling factor and $\Lambda$ is the ultraviolet (UV) cutoff which, in our model, is determined by the inverse lattice spacing $1/a$.
To obtain the effective action $S_\eff$ for the slow fields, one integrates out the fast components. With the shorthand notation $\mathscr{D}_>:=\mathscr{D}[\psi_{c/q>}^\pc,\psi_{c/q>}^*]$,
\begin{gather}
	e^{-S_\eff}:=\int \mathscr{D}_>\, e^{-S_R}
	=\int\mathscr{D}_>\, e^{-S_{G<}}e^{-S_{G>}}e^{-S_u}
	=e^{-S_{G<}}\,Z_{G>}\bra\, e^{-S_u}\ket_>\\
	\text{with}\quad
	Z_{G>}:=\int\D_>\, e^{-S_{G>}}\quad \text{and }\quad
	\bra O\ket_>:=\frac{1}{Z_{G>}}\int\D_>\, e^{-S_{G>}}\, O.
\end{gather}
Note that the integrand of the Gaussian action \eqref{eq:SR-G} only couples fields of equal momenta such that it can be easily split into the components $S_{G<}=\int_{\vx,t} [ \psi_{c<}^*\partial_t\psi_{q<}^\pc\dots]$ and $S_{G>}=\int_{\vx,t} [ \psi_{c>}^*\partial_t\psi_{q>}^\pc\dots]$ for the slow and fast fields, respectively. Also, when an observable $O$ is a function of slow fields only, then $\bra O\ket_>=O$. If $O$ is a function of fast fields only, then $\bra O\ket_>=\bra O\ket$ is a constant.
Expanding $S_\eff$ up to second order in $u$ and discarding the irrelevant constant term, we have
\begin{align}\nonumber
	S_\eff
	&\textstyle
	= S_{G<}-\ln \big\bra 1 - S_u+\frac{1}{2} S^2_u + \mc{O}(u^3) \big\ket_> -\ln Z_{G>}\\
	&=S_{G<}  +\bra S_u\ket_>-\frac{1}{2}\bra S_u^2\ket_{>,\text{connect}} + \mc{O}(u^3)
	\quad\text{with}\quad
	\bra S_u^2\ket_{>,\text{connect}}:= \bra S_u^2\ket_>-\bra S_u\ket_>^2.
	\label{eq:app03-Seff}
\end{align}

Using Wick's theorem \cite{Wick1950-80,Negele1988} and the free Green's function \eqref{eq:gf-appx}, $\bra S_u\ket_>$ evaluates to
\begin{align}
	\bra S_u\ket_>
	&= \frac{1}{2} \int_{\vx,t} \left\bra u \psi_c^2(\vx,t)\psi_c^*(\vx,t)\psi_q^*(\vx,t) - \text{c.c.}\right\ket_> \nonumber\\
	&= S_{u<}+  \int_{\vx,t}\left[ u\psi_{c<}^\pc(\vx,t)\psi_{q<}^*(\vx,t)-\text{c.c.}\right]
	   \int^{\Lambda}_{\Lambda/b}\frac{\ud^dk}{(2\pi)^d}\int\frac{\ud\omega}{2\pi}\,\bra\phi_c^\pc(\vk,\omega)\phi_c^*(\vk,\omega)\ket +const. \nonumber \\
	&= S_{u<}+  \int_{\vx,t}\left[u \psi_{c<}^\pc(\vx,t)\psi_{q<}^*(\vx,t)-\text{c.c.}\right]
	   \int^{\Lambda}_{\Lambda/b}\frac{\ud^dk}{(2\pi)^d}\int\frac{\ud\omega}{2\pi}\,
	   \frac{t_2}{|-\mri\omega+\mri J k^2+t_1+K k^2|^2} +const.  \nonumber \\
	&= S_{u<}+  \int_{\vx,t}\left[u \psi_{c<}^\pc(\vx,t)\psi_{q<}^*(\vx,t)-\text{c.c.} \right]
	   \int^{\Lambda}_{\Lambda/b}\frac{\ud^dk}{(2\pi)^d}\,
	   \frac{t_2}{2K k^2+t_1+t_1^*} +const.
	\label{eq:app03-SU}
\end{align}
In the step to line two, all fields have been split into slow and fast components, $\psi_{c/q}=\psi_{c/q<}+\psi_{c/q>}$, and Wick's theorem is applied. The terms with four slow fields constitute $S_{u<}$, and those with four fast fields yield the unimportant constant $\bra S_{u>}\ket$. The terms with an odd number of fast fields give zero. The remaining terms are products of two slow and two fast fields. The frequency integral over the contractions $\bra\psi_{c>}^\pc\psi_{q>}^*\ket$ and $\bra\psi_{q>}^\pc\psi_{c>}^*\ket$ give zero as, according to Eq.~\eqref{eq:gf-appx}, they just have a single pole (either on the upper or lower half of the complex $\omega$ plane) such that
\begin{align}
	\int\frac{\ud\omega}{2\pi}\,\bra\phi_c(\vk,\omega)\phi_q^*(\vk,\omega)\ket=\int\frac{\ud\omega}{2\pi}\,\bra\phi_q(\vk,\omega)\phi_c^*(\vk,\omega)\ket=0.
\end{align}
With $\bra\psi_{q>}^\pc\psi_{q>}^*\ket=0$, only the contraction $\bra\psi_{c>}^\pc\psi_{c>}^*\ket$ remains, giving line two of Eq.~\eqref{eq:app03-SU}. For the integral in line three, we can assume that $|t_1|$ is small compared to $K \Lambda^2$ since we are assessing the vicinity of the Gaussian fixed point $t_1=u=0$. This determines the locations of poles in the complex $\omega$ plane and, applying the residue theorem, we arrive at line four.

The evaluation of $\bra S_u^2\ket_{>,\text{connect}}$ is more involved but still straightforward. Discarding RG irrelevant terms and constants, and expanding the result up to leading order in $t_1$, its contribution to $S_\eff$ reads
\begin{align} \nonumber
	-\frac{1}{2}\bra S_u^2\ket_{>,\text{connect}} =
	&\int^{\Lambda}_{\Lambda/b}\frac{\ud^dk}{(2\pi)^d}\,
	\left(
	   \frac{-t_2 u^2}{4K^2 k^4}+\frac{-t_2 u^2}{8K (K+\mri J)k^4}
	   +\frac{-t_2 u u^*}{4K^2 k^4}
	   \right)
	   \int_{\vx,t}\psi_{c<}^2\psi_{c<}^*\psi_{q<}^*
	   \ - \,\, \text{c.c.} \\
	   &+ \int_{\vx,t}\Big( p_1(u) \psi_{q<}^* \psi_{c<}^\pc
	   + p_2(u) \psi_{q<}^*\partial_t\psi_{c<}^\pc
	   + p_3(u) \psi_{q<}^*\nabla^2\psi_{c<}^\pc \ - \,\, \text{c.c.} \Big)	   
	\label{eq:app03-SU2}
\end{align}
Here, we have only written out the terms where exactly four or two of the eight field variables in the product $S_u^2$ are slow and the others are fast: Terms with an odd number of fast fields give zero, the term with eight fast fields gives an unimportant constant, the term with eight slow fields does not contribute because of being unconnected, and the terms with two fast fields give RG irrelevant sixth-order contributions for $S_\eff$. For the terms with exactly four fast fields, leading to the first line in Eq.~\eqref{eq:app03-SU2}, at least one of the fast fields has to be $\psi_q$ or $\psi^*_q$ to get an RG relevant contribution. On the other hand, involving both the $\psi_q$ and the $\psi^*_q$ field in the contraction leads to a vanishing frequency integral, where all poles of the integrand are either in the upper or lower half of the complex plane. This is consistent with the fact that all terms generated in the RG are necessarily at least linear in $\psi_q$ or $\psi_q^*$ due to the conservation of probability \cite{Kamenev2023,Sieberer2016-79}. Hence, we have two contractions, where one $\psi_q$ or $\psi^*_q$ is involved.
The second line in Eq.~\eqref{eq:app03-SU2} arises from the terms with six fast fields, which give contributions of order $u^2$ for the flow of $t_1$, $K+\mri J$, and the $\partial_t$ term. There will be no correction to $t_2$, i.e., no correction to the $\psi_q\psi_q^*$ term due to the momentum conservation in the Wick contraction. The unspecified prefactors $p_i(u)=\alpha_i u^2+\beta_i uu^*$ are quadratic in $u$.

The effective action \eqref{eq:app03-Seff} for the slow fields then reads
\begin{align}\nonumber
	S_\eff=\int_{\vx,t}\Bigg[\, 
	 & \Big(\big(1+p_2(u)\big)\psi_{q<}^*\partial_t\psi_{c<}^\pc
	   -\big(K+\mri J-p_3(u)\big)\psi_{q<}^*\nabla^2\psi_{c<}^\pc
	   - c.c.\Big) + t_2\psi_{q<}^*\psi_{q<}^\pc \\ \nonumber
	 & +\Bigg(t_1+ \int^{\Lambda}_{\Lambda/b}\frac{\ud^dk}{(2\pi)^d}\,\frac{t_2u}{2K k^2+t_1+t_1^*}+p_1(u)\Bigg)  \psi_{q<}^* \psi_{c<}^\pc
	     \,\, - \,\,\text{c.c.}\\
	 & +\Bigg( \frac{u}{2} -\int^{\Lambda}_{\Lambda/b}\frac{\ud^dk}{(2\pi)^d}\,
	    \Big( \frac{-t_2 u^2}{4K^2 k^4}+\frac{-t_2 u^2}{8K (K+\mri J)k^4}
	         +\frac{-t_2 u u^*}{4K^2 k^4} \Big)\Bigg)
	     \psi_{c<}^2\psi_{c<}^*\psi_{q<}^*\,\, - \,\,\text{c.c.} \,\Bigg] + \mc{O}(u^3).
\end{align}
So, after integration over the fast fields, the effective action has the same structure as the original action \eqref{eq:SR} with the coupling coefficients replaced by
\begin{subequations}
\begin{align}
	t_1 &\to t_1+ \int^{\Lambda}_{\Lambda/b}\frac{\ud^dk}{(2\pi)^d}\,\frac{t_2u}{2K k^2+t_1+t_1^*} + \mc{O}(u^2),\\
	t_2 &\to t_2 +  \mc{O}(u^3),\quad
	K+\mri J \to K+\mri J +  \mc{O}(u^2),\quad
	\partial_t \to \big(1+\mc{O}(u^2)\big)\partial_t,\\
	\frac{u}{2}   &\to  \frac{u}{2} -\int^{\Lambda}_{\Lambda/b} \Big(
	   \frac{t_2 u^2}{4K^2 k^4}+\frac{t_2 u^2}{8K (K+\mri J)k^4}
	   +\frac{t_2 u u^*}{4K^2 k^4}
	   \Big)+\mc{O}(u^3).
\end{align}
\end{subequations}

The next step in the RG procedure is to rescale momenta and frequencies, and to renormalize the fields such that the leading-order unperturbed terms (the derivative terms and the $t_2$ term) are invariant. Here, we have three anomalous dimensions associated with $\partial_t$, $K+\mri J$, and $t_2$ as well as the three degrees of freedom in the renormalization of $\omega$, $\psi_{c}$, and $\psi_{q}$ to make the leading-order terms invariant. Up to $\mc{O}(u)$, the corresponding scaling dimensions agree with the canonical scaling dimensions (cf.\ Sec.~\ref{sec:scaling}),
\begin{equation}
	k\to bk,\quad \omega\to b^{2+\mc{O}(u^2)}\omega,\quad \psi_{c<}\to b^{(d-2)/2+\mc{O}(u^2)}\psi_{c}, \quad \psi_{q<}\to b^{(d+2)/2+\mc{O}(u^2)}\psi_{q}.
\end{equation}
In this way, the original UV cutoff is restored ($\Lambda/b\to\Lambda$), and we can read off the renormalized coefficients. In $d=4-\epsilon$ spatial dimensions,
\begin{subequations}\label{eq:app03-ttu}
\begin{align}
	t_1(b)&= b^2\left[ t_1+ \int^{\Lambda}_{\Lambda/b}\frac{\ud^dk}{(2\pi)^d}\,\frac{t_2u}{2K k^2+t_1+t_1^*}+\mc{O}(u^2)\right] ,\qquad
	t_2 (b)= t_2 ,\quad
	J (b)= J ,\quad
	K (b)= K, \\
	u(b)&=b^{\epsilon}\left[ u-\int^{\Lambda}_{\Lambda/b}\frac{\ud^dk}{(2\pi)^d}\, \Big(
	   \frac{t_2 u^2}{2K^2 k^4}+\frac{t_2 u^2}{4K (K+\mri J)k^4}
	   +\frac{t_2 u u^*}{2K^2 k^4}
	   +\mc{O}(u^3) \Big)\right].
\end{align}
\end{subequations}
Now, let us choose the rescaling factor as $b=1+\ud\ell$ with an infinitesimal $\ud\ell$ such that $t_1(b)$ and $u(b)$ read
\begin{subequations}
\begin{align}
	t_1(\ud\ell)& =(1+2\,\ud\ell)\left[ t_1(0)+\frac{t_2(0)u(0)}{2K(0)\Lambda^2+t_1(0)+t_1^*(0)}\Lambda^{d}\mc{S}_d \,\ud\ell +\mc{O}(u^2)\right],\\
	u(\ud\ell)&=(1+\epsilon\, \ud\ell)\left[
	   u(0)-\Big(\frac{t_2(0)u^2(0)}{2K^2(0)}
	            +\frac{t_2(0)u^2(0)}{4K(0)(K(0)+\mri J(0))}+\frac{t_2(0)u(0)u^*(0)}{2K^2(0)}
	\Big)\Lambda^{d-4}\mc{S}_d \,\ud\ell +\mc{O}(u^3) \right]
\end{align}
\end{subequations}
with the phase-space factor $\mc{S}_d:=2/[(4\pi)^{d/2}\Gamma(d/2)]$ resulting from the surface integral over the unit sphere. Lastly, setting $\Lambda=1$, we obtain the RG flow equations
\begin{subequations}\label{eq:app03-flow}
\begin{align}
	\frac{\ud t_1}{\ud\ell}&=2t_1+\frac{\mc{S}_d t_2}{2K+t_1+t_1^*}\, u+\mc{O}(u^2)\quad\text{and}
	\label{eq:app03-flow-t1}\\
	\frac{\ud u}  {\ud\ell}&=\epsilon u-\mc{S}_d\left(\frac{t_2u^2}{2K^2}+\frac{t_2u^2}{4K(K+\mri J)}+\frac{t_2 uu^*}{2K^2}\right)
	                          +\mc{O}(u^3),
	\label{eq:app03-flow-u}
\end{align}
\end{subequations}
which are discussed in Sec.~\ref{sec:below} of the main text.

For $J=0$, both $t_1$ and $u$ remain real during the flow, and we can solve the flow equations \eqref{eq:app03-flow}. For general $\epsilon=d_c-d$, the trajectories involve the hypergeometric function. For three-dimensional space ($\epsilon=1$), one finds the simple form
\begin{equation}
	t_1(u,C)=u\,\frac{C u - 8\mc{S}_d t_2K^3-10(\mc{S}_d t_2)^2K \, u\ln u}{(5\mc{S}_d t_2 u-4K^2)^2} + \mc{O}(u^2)
	\quad\text{with}\quad C\in\RR,
\end{equation}
and the critical manifold, separating the two phases of the system, is the flow trajectory $t_1(u,\tilde{C})$ with $\tilde{C}=10(\mc{S}_d t_2)^2K[1+\ln(4K^2/5\mc{S}_d t_2)]$. Figure~\ref{fig:RG_flow} shows the flow diagram for this case.

\twocolumngrid
\bibliographystyle{prsty.tb.title}

\end{document}